\documentclass[12pt,preprint]{aastex}
\newcommand{\hst}{{\sl HST}}

\slugcomment{ApJL accepted on 17 March 2005}
\shorttitle{The the WDs Cooling sequence in NGC~6791}
\shortauthors{Bedin et al.}
\begin{document}
\def\subr #1{_{{\rm #1}}}
\title{THE WHITE DWARF COOLING  SEQUENCE IN NGC\ 6791}

\author{ Luigi R.\ Bedin\altaffilmark{1,2},  
         Maurizio Salaris\altaffilmark{3},  
         Giampaolo Piotto\altaffilmark{2}, 
         Ivan R.\ King\altaffilmark{4}, 
         Jay Anderson\altaffilmark{5}, 
         Santi Cassisi\altaffilmark{6}, 
         \and 
	 Yazan Momany\altaffilmark{2} 
         }

\altaffiltext{1}{European             Southern            Observatory,
Karl-Schwarzschild-Str.\ 2, D-85748 Garching, Germany; lbedin@eso.org}

\altaffiltext{2}{Dipartimento  di Astronomia, Universit\`a  di Padova,
Vicolo dell'Osservatorio 2, I-35122 Padova, Italy; piotto-momany@pd.astro.it}

\altaffiltext{3}{Astrophysics Research Institute, Liverpool John Moores
Univ., 12 Quays House, Birkenhead, CH41 1LD, UK; ms@astro.livjm.ac.uk}

\altaffiltext{4}{Department of Astronomy, University of Washington,
Box 351580, Seattle, WA 98195-1580; king@astro.washington.edu}

\altaffiltext{5}{Department of Physics and Astronomy, Mail Stop 108,
Rice University, 6100 Main Street, Houston, TX 77005;
jay@eeyore.rice.edu}

\altaffiltext{6}{INAF-Osservatorio Astronomico di Collurania,
via M. Maggini, 64100 Teramo, Italy;
cassisi@te.astro.it}

\begin{abstract}
In the  old, populous, and metal-rich  open cluster NGC\  6791 we have
used deep  \hst/ACS images to  track the white dwarf  cooling sequence
down  to  $m_{\rm   F606W}\simeq28.5$.   The  white  dwarf  luminosity
function shows a well defined peak at $m_{\rm F606W}\simeq27.4$, and a
bending to  the blue  in the color--magnitude  diagram.  If  this peak
corresponds  to  the end  of  the  white  dwarf cooling  sequence  the
comparison with theoretical isochrones provides a cluster age estimate
of $\sim 2.4$ Gyr, in sharp contrast with the age of 8--9 Gyr inferred
from the main-sequence turn-off.  If the end is at fainter magnitudes,
the  peak at  $m_{\rm F606W}\simeq27.4$  is even  more  enigmatic.  We
discuss possible causes, none of them very convincing.
\end{abstract}


%
\section{Introduction}
%

NGC\ 6791 is  one of the richest open clusters, with  age $>8$ Gyr and
([Fe/H]  $\sim$ +0.4).  Because  of these  unusual properties,  it has
been the target of several studies, the most recent by Carney, Lee, \&
Dodson (2005).
 
NGC\ 6791 is close enough that  \hst/ACS imaging can hope to reach the
bottom  of  the  white  dwarf  (WD) cooling  sequence,  and  therefore
estimate  its age  independently  of the  main-sequence (MS)  turn-off
(TO).  For  clusters older than  3 Gyr this  kind of age  estimate has
previously been attempted only in the $\sim$4-Gyr-old open cluster M67
(Richer et  al.\ 1998),  and the globular  cluster M4 (Hansen  et al.\
2002, 2004).  In  the latter case the end of  the cooling sequence was
not detected, and the age estimate  was based on modeling the shape of
the observed part of the WD luminosity function (LF).
 
The data on which this letter  is based were originally taken to reach
the bottom of  the MS in NGC\ 6791,  but when we first saw  the CMD it
was clear that the WD cooling sequence (WDCS) was equally exciting.  A
companion paper will describe  the color--magnitude diagram and the MS
LF of  NGC 6791,  and compare the  observations with models.   Here we
focus on the WD sequence.

%
\section{Observations and Measurements}
%

The  data were  taken on  July 16--17  2003, with  \hst's  ACS/WFC (GO
9815), and  consist of 6 long  ($\sim$1150s) and 3  short exposures of
$\sim$30s through  each of  the filters F606W  and F814W.   A detailed
discussion  of  the   data  reduction,  photometric  calibration,  and
artificial star  tests is presented in our  companion paper.  Briefly,
the images in  each filter were corrected for  distortion according to
Anderson  (2002),  and combined  into  a  stack  in which  pixels  are
sub-sampled  by  a  factor   of  two  to  remove  undersampling.   The
photometry  was  performed with  DAOPHOT  (Stetson 1987).   Artificial
stars were  added in order  to determine the completeness  curve.  The
calibration to the ACS Vega-mag  system was done following the recipes
in Bedin et al.\ (2005).
 
Figure 1  summarizes our observational results.  Panel  $a)$ shows the
color--magnitude  diagram   (CMD),  while  panel   $b)$  displays  the
completeness-corrected WD  LF.  There  is a clear  peak of  density at
$m_{\rm  F606W}\sim27.4$, after  which the  LF drops,  but  stays well
above  zero down  to $m_{\rm  F606W}\sim28.2$, where  our completeness
becomes $\le  50$\%.  The peak in  the LF coincides with  a bending in
the CMD of  the WDCS.  At $m_{\rm F606W}\sim28.2$ there  seems to be a
second peak,  but its significance remains to  be investigated.  Panel
$c)$ shows  the input  (continuous line) and  the output CMD  from the
artificial-star  experiments, and panel  $d)$ the  corresponding joint
(two filters) completeness curve for WDs.  Panels $e)$ and $f)$ show a
blow-up  of  the CMD  focused  on  the  observed and  artificial  WDs,
respectively.
 
It  is evident from  panels $b)$  and $d)$  that at  the level  of the
density peak  in the WD LF  the completeness is  still $\sim85$\%, and
therefore the peak and the bending to the blue are real.
At $\ell\sim70^\circ$, $b\sim+11^\circ$  we expect the contribution of
field WDs in our small field to be negligible, and because we rejected
all non-point-like sources by  visual inspection, we expect the number
of contaminating galaxies also to be negligible.

%
\section{Comparison with theory}
%

In Fig.~2,  DA WD isochrones for  ages of 2, 4,  6, 8, and  10 Gyr are
compared with the observed WD  sequence.  The isochrones came from the
WD models by Salaris et al.\ (2000), transformed into the ACS Vega-mag
system  using   the  most  up-to-date   on-orbit  transmission  curves
(Sirianni et al.\ 2005), and color transformations generously provided
by  P.\ Bergeron,  based on  the model  atmospheres used  in Bergeron,
Leggett, \& Ruiz  (2001).  In addition to the  cooling models, we have
adopted  the empirical relation  between WD  mass and  progenitor mass
initial-final   mass  relation,  IFMR)   by  Weidemann   (2000).   The
progenitor lifetimes come from the Pietrinferni et al.\ (2004) grid of
stellar models.
 
In the WD model-fitting we adopted the distance and reddening found in
our companion  paper, where we showed that  the best fit to  the MS TO
and  subgiant branch  (reproduced  also in  Fig.~2)  is obtained  with
[Fe/H] = 0.26,  an apparent modulus $(m-M)_{\rm F606W}=  13.44$, and a
reddening $E(m_{\rm F606W}-m_{\rm F814W})= 0.14$.  These correspond to
$(m-M)_V$=13.50 and $E(B-V)$=0.15 in the Johnson band-passes (Bedin et
al.\ 2005).   Notably, the age derived  from the fitting of  the TO is
between 8  and 9  Gyr, as  discussed in more  detail in  our companion
paper.   All these  values  are  in broad  agreement  with the  recent
analysis by Carney et al.\ (2005).

The brighter parts of our  WD isochrones are populated by objects with
mass  of  $\sim$0.54 $m_{\odot}$---i.e.,  WDs  produced  by the  stars
presently at  the asymptotic giant  branch (AGB) tip.   The isochrones
reproduce  well the  main body  of  the observed  WD sequence  between
$m_{\rm F606W}\sim 24$  and $m_{\rm F606W}\sim 26$.  A  few objects in
this magnitude  range lie  to the red  side of  the main locus  of the
WDCS, most probably a mixture  of field contamination, He-core WDs and
possibly unresolved binary WD-WD systems.
 
The bottom of  the WD isochrones shows  a turn to the blue  due to the
appearance  of larger  WD masses,  from  the more  massive stars  that
evolved off the AGB early  in the cluster's history. The brightness of
this turn to the blue is a function of cluster age (see, e.g., Salaris
et al.\ 2000, Richer et al.\ 2000).  It is clear from Fig.~2 that even
the bottom of the 6 Gyr WD isochrone is at a brightness where the peak
in the WD population has already dropped.
 
Figure 2  highlights {\it a  potentially serious mismatch  between the
MSTO  and WD  ages for  NGC\ 6791}.   For comparison  we also  show in
Fig.~2  the  location  of  the   sample  of  cool  WDs  in  the  solar
neighborhood  (open  squares) analyzed  by  Bergeron  et al.\  (2001),
translated into the same  ACS Vega-mag system with the transformations
adopted  in this  paper.  As  expected,  the local  WDs reach  fainter
magnitudes than  the bulk  of the cluster  WDCS, down to  a brightness
consistent with our 8 Gyr isochrone.
 
To investigate further this age  mismatch, we consider again the LF of
the cluster WDCS  (see Fig.~3 {\it top} panel),  with its well-defined
peak at $m_{\rm  F606W}\sim$ 27.4.  From a theoretical  point of view,
one can determine  a LF for a given age  starting from the appropriate
WD  isochrone and  adopting an  initial  mass function  (IMF) for  the
progenitors (see, e.g., Richer et al.\ 2000); all theoretical LFs show
a peak at the faint end  that moves toward fainter luminosities as the
cluster  age  increases, reflecting  the  behavior  of the  underlying
isochrones.  This peak is caused by the ``piling up'' at the bottom of
the WD  isochrone of WDs  of all masses,  due to their  finite cooling
age.   If one uses  a power-law  IMF ($dn/dm\propto  m^{-(1+x)}$), the
exact value of  the exponent $x$ ($x=1.35$ for  Salpeter IMF) does not
affect the position  of the LF peak.  The position of  the peak of the
observed WDCS can  be reproduced only by assuming  an age below 3~Gyr,
in  clear disagreement with  the estimated  TO age  of about  8--9 Gyr
($\sim$1.5~mag  fainter).   A  sub-population  of stars  with  age  of
$\sim$3 Gyr  can be excluded  by the fact  that we do not  observe the
corresponding TO.
 
Further  evidence that we  are facing  a serious  problem is  given in
Fig.~3 ({\it middle}  and {\it bottom} panels), which  compares the WD
LF of  the solar-metallicity  open cluster M67  (Richer et  al.\ 1998)
with our NGC 6791 WD LF.   The NGC\ 6791 WD data have been transformed
into  the   Johnson-Cousins  system  from  the  ACS   one,  using  the
transformations by  Sirianni et al.\ (2005), which  should be accurate
to a  few hundredths  of a magnitude.   The distance moduli  have been
subtracted from both LFs (we used values in Percival \& Salaris 2003),
and  the NGC\ 6791  LF has  been rebinned  in order  to match  the M67
binning.  The  M67 LF also shows  a peak, which  most probably denotes
the bottom of  the DA WD sequence.  Surprisingly  enough, the peaks in
the two clusters  are at the same absolute magnitude.   But the TO age
of M67 is
only $\sim$4 Gyr  (see, e.g., Salaris, Weiss \&  Percival 2004), i.e.,
about half the NGC\ 6791 age.  With our WD isochrones, using lifetimes
for solar-metallicity progenitors, we  can reproduce the observed peak
in the M67  WD LF adopting an  age of 3.5 Gyr, consistent  with its TO
age. The same  result is obtained by Richer et  al.\ (1998, 2000) with
their own WD  isochrones.  The fact that the TO and  WD ages agree for
M67 makes the disagreement in NGC\ 6791 even more puzzling.

%
\section{Discussion}
%

As shown in  the previous paragraph, the nature of the  the WD LF peak
at  $m_{\rm F606W}\simeq27.4$  is enigmatic.   We are  faced  with two
possibilities:\ 1) either the peak  represents the bottom of the WDCS,
and in this  case we need to  understand why the WD cooling  age is in
sharp contrast with the age inferred from the MS TO; or 2) the peak is
not due  to the classical  DA WDs.  In  the following we  will further
discuss the consequences of these two hypotheses.
 
Let us  first examine  the possibility that  the observed  peak indeed
represents the bottom  of the WDCS of NGC\ 6791.   A wrong cooling age
could simply come from having adopted a wrong distance modulus, but an
apparent distance modulus smaller by 1.5 mag would imply a TO age well
above 20  Gyr.  On the other hand,  we cannot exclude that  the TO age
could be  wrong because the  adopted metallicity is wrong,  though the
most  likely  metallicity  interval  (Taylor 2001)  cannot  solve  the
discrepancy.

Another possibility  is an  IFMR radically different  from the  one we
used.  However, as we have verified with various numerical tests, none
of the proposed IFMR in Fig.\  11 of Hansen et al.\ (2004) helps solve
the problem.  As noted by the referee (H.\ Richer), the WD LF shows an
apparent gap  (of low  statistical significance, however),  at $M_{\rm
F606W}  \simeq12.2$,  which  could  correspond to  a  similar  feature
identified  in  M4  at   $M_{\rm  F606W}\simeq13.2$  (Hansen  et  al.\
2004)---a full  magnitude fainter. Is  this further evidence  that the
cooling models  specific for NGC\  6791 fail?  In the  magnitude range
covered  by the  bottom  of the  observed  WD sequence  all modern  WD
cooling models---including  the Salaris et al.\ (2000)  models we have
adopted  here---provide   very  similar  cooling   times  (see,  e.g.,
Prada-Moroni \& Straniero 2002).

Another possibility  of explaining  this age mismatch  is to  invoke a
wrong CO  chemical stratification in the  WD core of  the models (that
underestimates  the  energy  available  during  the  cooling  and  the
crystallization process) and/or a wrong atmospheric H-layer thickness.
However,  the most  recent  determination of  the $\rm  C^{12}(\alpha,
\gamma)O^{16}$ reaction  rate would not  solve the problem,  nor would
the largest possible value of $\Delta m_{\rm H}$ for the H-layer.

The CMD of Fig.~1 clearly shows  that NGC\ 6791 has a large population
of binaries.   It might be that  the LF peak  is made of DA  WDs whose
progenitors  were  produced by  merging  (or  stripping) in  binaries,
possibly during their MS phase.  If these mergers occur about 2--3~Gyr
after the birth  of the cluster and the  resulting objects have masses
of $\sim$1.5~$m_{\odot}$,  their WD progeny would  populate the region
of the LF  peak for a cluster age of about  8~Gyr.  The counterpart of
these merged progenitors might  be the blue stragglers (BSs).  Davies,
Piotto, \& De Angeli (2004), based on observational evidence by Piotto
et  al.\ (2004),  show that  the number  of BSs  coming  from evolving
binaries should have been much larger in the first few Gyr of the life
of  the cluster.   We  also  note that  the  HB of  NGC\  6791 has  an
anomalous morphology (Yong  et al.\ 2000). The origin  of this anomaly
is not at all clear, and it  is even less clear how it propagates into
the WD regime.

Let us  now explore the case  that the LF peak  is not due  to DA WDs.
WDs with a degenerate He-core and masses of 0.3--0.4 $m_{\odot}$, from
progenitors of $\sim$1.3--1.5 $m_{\odot}$,  can populate the region of
the observed peak if we adopt  a cluster age of $\sim$8 Gyr, according
to  the He-core WD  models by  Serenelli et  al.\ (2001).   However, a
substantial contamination by He-core  WDs can be excluded because, due
to their lower mass, they are much redder than the bulk of the WDCS we
observe.  We  explicitly verified this point by  transforming into the
ACS Vega-mag photometric system the  He-core WD models by Serenelli et
al.\  (2001) for  masses of  0.32 and  0.41 $m_{\odot}$  (see Fig.~2).
Moreover, using as  normalizing factor the number of  stars on the MS,
and adopting a Salpeter IMF, we  estimated that the total number of CO
WDs in our  field should be $\sim$430, which  compares reasonably well
with the $\sim$600  objects observed.  Though ignorance of  the IMF at
higher  masses,  and  mass   segregation  effects,  make  this  number
extremely  uncertain,  there  seems  little  room  for  a  substantial
population  of  He-core  objects.    For  DA  WD  luminosities  around
$\log(L/L_{\odot})\sim -4.0$, non-DA objects are slightly fainter than
DAs ($\sim$0.5 mag,  but the exact value depends on  the WD mass), and
the  difference  increases  fast  with decreasing  luminosity.   These
objects may  contribute to the possible second  peak around $\sim$28.2
but, again,  only for ages  around 3~Gyr.  Also, the  possibility that
the peak comes from a cluster  of far, unresolved blue galaxies can be
excluded,  as the  spatial distribution  of  the observed  WDs is  not
significantly different from that of the cluster MS stars.
 
We also note that the WD LF  of Fig.~1b never drops to zero. This fact
could indicate that we have not  reached the end of the WDCS.  In this
case the presence of a peak in the WDCS may also indicate the onset of
an  unknown  energy  release  at  this specific  age  and  metallicity
(sedimentation/separation  of CO,  pycnonuclear reaction,  or whatever
can cause a temporary slowdown in the cooling).

Finally,  it is  worth  remarking that  NGC\  6791 is  a cluster  with
several peculiarities (high binary  fraction, anomalous HB, very metal
rich,  dynamically  old, etc.)   that  may  have  altered the  stellar
populations in  the cluster,  and hence  it is too  early to  call for
major  revisions in  the WD  formation  and evolution  models.  It  is
surely of great interest to  extend the present investigation at least
down to the  magnitude where we expect  to see the peak of  a 8--9 Gyr
WDCS.

\acknowledgments

We are grateful to Pierre  Bergeron and to Alvio Renzini.  I.R.K.\ and
J.A.\  acknowledge support by  STScI grant  GO-9815.  S.C.,  G.P., and
Y.M.\ acknowledge support by MIUR under PRIN2003.

\clearpage

\begin{figure}
\epsscale{1.00}
\plotone{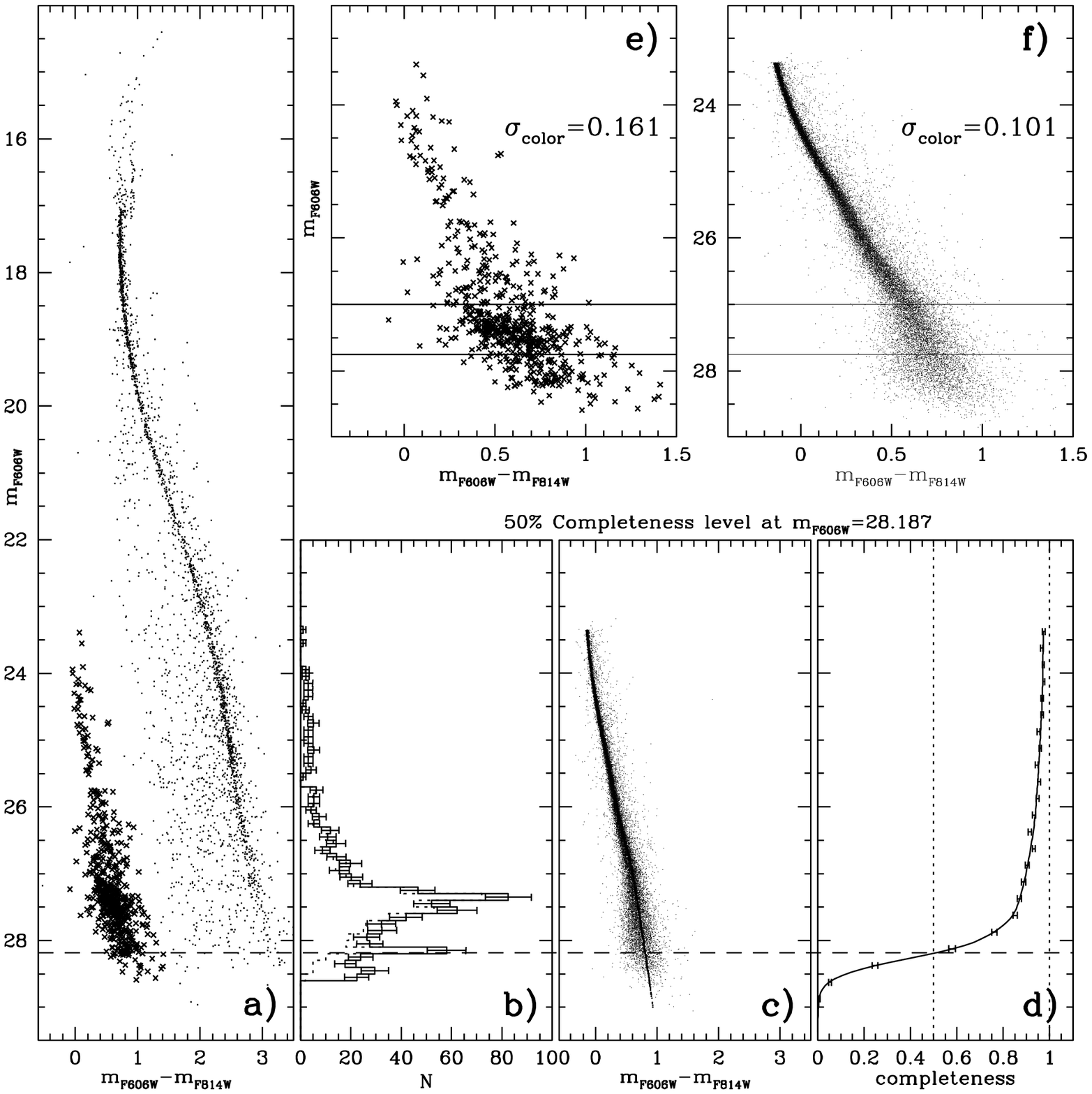}
\caption{The  CMD and  WD  LF  of NGC\  6791  and the  artificial-star
results (see  text).  The broadening in  color of the  observed CMD at
the  level of the  WD peak  is $\sim60\%$  larger than  for artificial
stars.}
\end{figure}

\begin{figure}
\epsscale{1.00}
\plotone{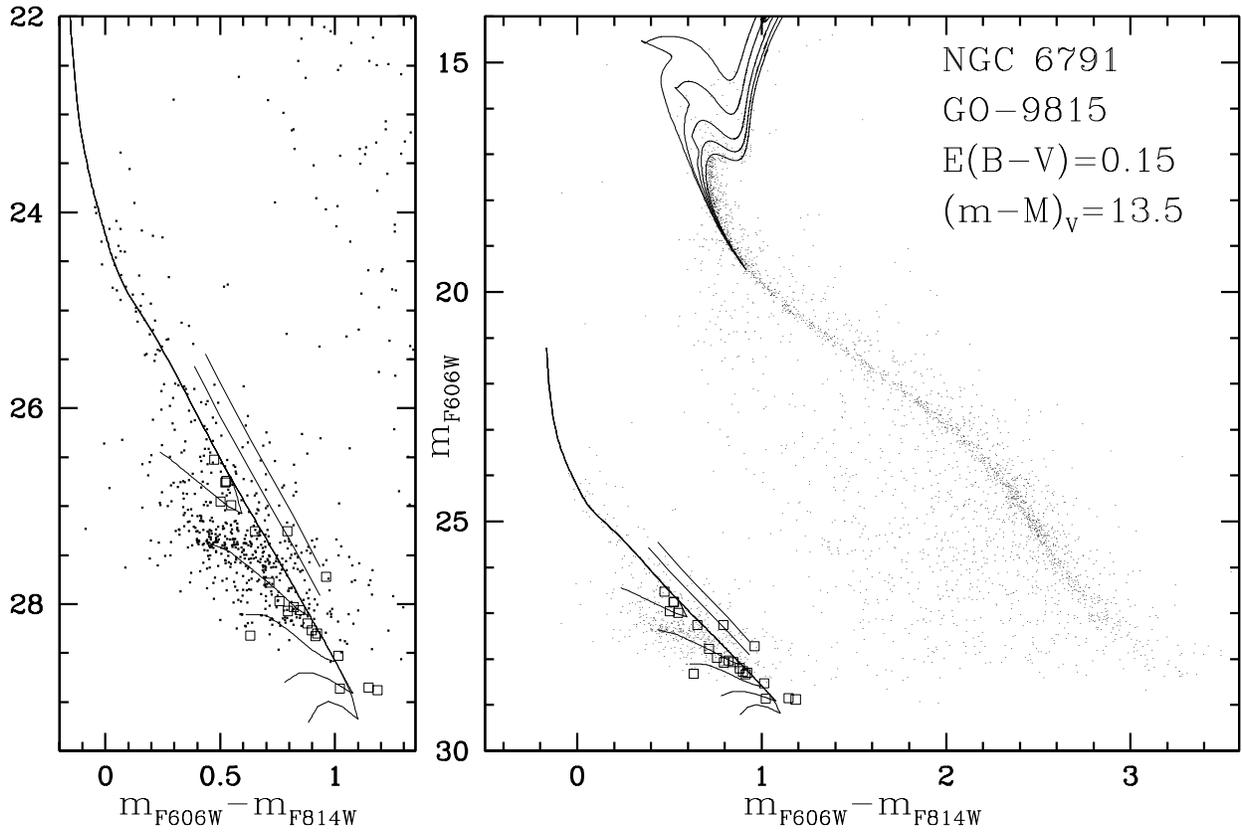}
\caption{The MS and the WDCS of NGC\ 6791, fitted with 2, 4, 6, 8, and
10 Gyr isochrones, with the  same distance modulus and reddening.  The
lines on  the red side of the  WD isochrones indicate the  loci of WDs
with pure helium cores of 0.32 and 0.41 $m_\odot$.}
\end{figure}

\begin{figure}
\epsscale{1.00}
\plotone{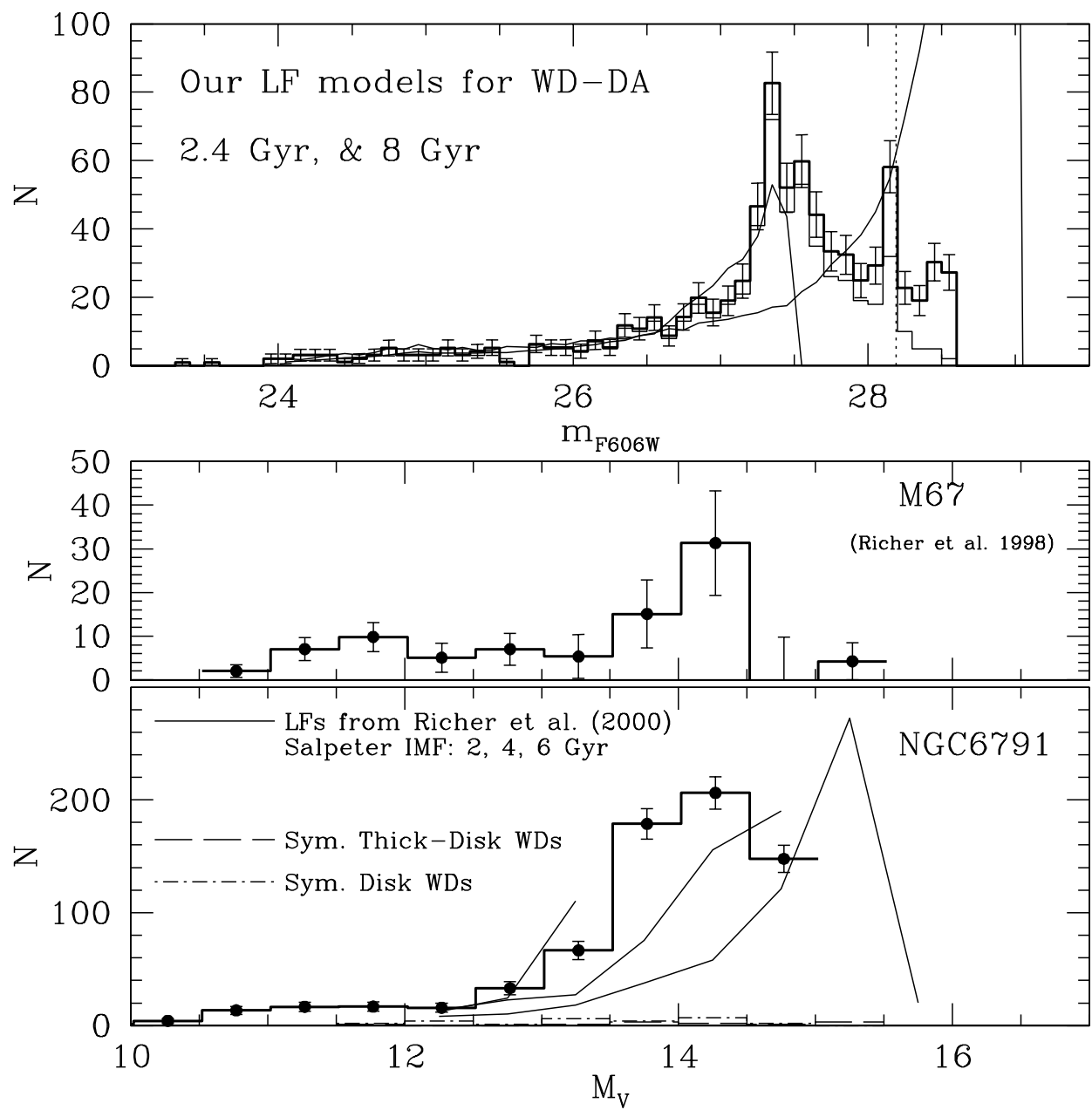}
\caption{ ({\it  Top} panel)  The completeness-corrected LF,  with the
theoretical LFs  for 2.4 and  8 Gyr overplotted.  The  vertical dotted
line is the  50\% completeness level.  The WD LF  of M67 ({\it middle}
panel) and that NGC\ 6791 ({\it  bottom} panel) have peaks at the same
absolute magnitude, corresponding to an age $<$4 Gyr.}
\end{figure}

\end{document}